# Thermal Conductivity Modeling of Monodispersed Microspheres using Discrete Element Method

Jian Zeng[*1], Ka Man Chung[*2], Xintong Zhang[*1], Sarath Adapa[1], Tianshi Feng[1], Yu Pei[1], Renkun Chen[#1, 2]

[1]Department of Mechanical and Aerospace Engineering, University of California, San Diego, La Jolla, California 92093, United States

[2]Material Science and Engineering Program, University of California, San Diego, La Jolla, California 92093, United States

*These authors contributed equally.

#Corresponding author: rkchen@ucsd.edu

**Abstract**

Particle beds are widely used in various systems and processes, such as particle heat exchangers, granular flow reactors, and additive manufacturing. Accurate modeling of thermal conductivity of particle beds and understanding of their heat transfer mechanisms are important. However, previous models were based on a simple cubic packing of particles which could not accurately represent the actual heat transfer processes under certain conditions. Here, we examine the effect of the packing structure on thermal conductivity of particle beds. We use monodispersed silica microspheres with average article sizes ranging from 23 to 330 µm as a model material. We employ a transient hot-wire technique to measure the thermal conductivity of the particle beds with packing density of 43 to 57% within a temperature range of room temperature to 500°C and under $N_2$ gaseous pressures of 20 to 760 Torr. We then use a discrete element method (DEM) to obtain the realistic packing structure of the particles, which is then fed into a finite-element model (FEM) to calculate the thermal conductivity, with the consideration of solid conduction, gas conduction, and radiation heat transfer. Our results show that the thermal conductivity model based on the more realistic random packing structure derived from the DEM shows better agreement with the experimental data compared to that based on the simple cubic packing structure. The combined DEM and FEM methodology can serve as a useful tool to predict effective thermal conductivity of particle beds and to quantify different heat transfer mechanisms under various conditions.

**I. Introduction**

Particle beds are useful for several current and emerging systems and processes, such as catalytic reactors [1], nuclear reactors [2,3], heat exchangers[4], thermal and thermochemical energy storage[5,6], thermal insulation[7], adsorption bed[8], and additive manufacturing[9]. Heat transfer performance of these systems is highly dependent on the effective thermal conductivity of the particle beds ($k_{eff}$) [10,11], and it is thus important to be able to model and even predict $k_{eff}$ to optimize the system performance. Despite continuous efforts in the modeling of $k_{eff}$ of particle beds with varying materials, sizes, and packing structures since the 1960s [12-18], an accurate and predictive model remains elusive due to the complexity of heat transfer in the particle bed. $k_{eff}$ represents the collective heat transfer performance from the solid conduction, gas conduction and radiation in a heterogeneous fluid and solid system. As such, it is affected by numerous factors, such as particle size, porosity, the intrinsic thermophysical properties of the solid and fluid as functions of temperature and pressure, the interaction between the solid surface and the stagnant fluid and the mechanical adhesion between the solid surface, making it challenging to accurately model $k_{eff}$ under a wide range of conditions.

Analytical models have been developed in the past few decades to capture $k_{eff}$ of packed particles. The simplest models include the in-series and in-parallel thermal resistances between the solid and fluid phases. The in-series model predicts the lower bound of $k_{eff}$ of a solid-fluid system while the in-parallel one represents the upper bound. More sophisticated models for particulate media are based on unit cell with both in-series and in-parallel thermal pathways[12-18]. For example, in the models developed by Yagi and Kunni, Kunii and Smith, and Masamune and Smith [14-16], there are generally three different heat transfer mechanisms that constitute the effective conduction pathways: the first is the heat transfer through the void fraction by gaseous conduction and radiation; the second mechanism is the solid-gas conduction where the solid and fluid layers are in series; and the third one is the heat transfer within the solid phase through the particle contacts. Their models, however, are limited to periodic packing structures which might not represent the realistic ones. On the other hand, the key parameters, for example, the particle contact fraction and effective solid-gas conduction pathway are empirically obtained based on experimental results. It might have oversimplified the solid-gas and solid contact pathways. Other types of unit cells have also been proposed to better capture the heat transfer mechanisms in packed particles[19,20]. For example, Hall and Martin [19] developed models for primitive cubic-packed array of spheres and for squared packed infinite cylinders. Although the models can include the low gas pressure and radiation effects, the models are limited to primitive cubic-packing structures. In addition, these models have not taken solid-solid heat conduction into consideration. The Zehner, Bauer, and Schlünder (ZBS) model is one of the widely used models, which considers packed particles within a cylindrical unit cell[20]. However, the ZBS model has limited accuracy under several conditions due to its simplified particle packing assumption. The analysis of the heat conduction through the finite solid-to-solid contact areas also requires the full knowledge of the elasticity of the solid particle. Thus, it is difficult to accurately predict $k_{eff}$ of packed particles by simply applying the ZBS model. Therefore, the existing analytical models are usually limited by the assumption of regular particle packing and simplified particle-to-particle contact in a unit cell.

Realizing the importance of accurate packing structure on the prediction of $k_{eff}$, several studies have been carried out to model the structure of randomly packed beds via the discrete element method (DEM) with great success on a few numbers of materials[21-24]. However, most of the prior studies were focused on large particles as the intended application was primary on nuclear reactors, e.g., 600 to 1000 μm diameter pebbles with standard deviations of ~ 150 μm[21,22,24,25]. It is still of fundamental interest to examine the impact of particle packing structure on a wider range of particle sizes and under various gaseous pressures. This is because the packing structure is expected to impact the solid-to-solid and the gas-solid heat conduction mechanisms, which are known to depend on particle size and gaseous pressure [4,26,27]. Therefore, examining the impact of the realistic random packing structure vs. the conventional periodic structure on $k_{eff}$ modeling could be more relevant to assess the suitability of DEM over a wide range of experimental conditions. In this work, we aim to carry out a systematic study on monodispersed spherical particles within a wide range of particle size, temperature and gaseous pressure to evaluate the random particle packing structure derived from a DEM model and to quantify the contribution of different heat transfer pathways. Although the effects of the parameters (particle size, temperature, pressure) have been investigated separately in several earlier studies[23,28,29], the limited range of values in each of these studies and the different experimental conditions across different studies still inhibited effective validation of the DEM model and quantification of contributions of different heat transfer mechanisms.

In this work, monodispersed silicon dioxide ($SiO_2$) microsphere particles of different sizes (~23 μm, ~67 μm and ~330 μm) were chosen as a model system to study the heat transfer mechanisms in packed particles. $k_{eff}$ of the $SiO_2$ microspheres was first experimentally determined by using a high-temperature transient hot-wire (THW) setup, in the temperature range of room temperature to ~500°C. Gaseous pressures of the particle beds were systematically varied to tune the contribution from the gas pathway. We then developed a model using DEM to generate the random packing structure of the particles, which is then incorporated into a COMSOL based finite element model (FEM) to predict $k_{eff}$. The DEM model can simulate a more realistic three-dimensional (3D) structure of the monodispersed particles with the input of relevant material properties. The $k_{eff}$ simulation results based on the packing structure obtained from the DEM model shows excellent agreements with the experimental results within the entire temperature and gaseous pressure ranges, whereas the model based on a simple cubic packing structure shows large discrepancies under several experimental conditions with the experimental results. Our work suggests that DEM can serve as a useful tool to predictively model thermal conductivity of particle beds with more realistic particle packing structure and leads to better understanding of various heat transfer pathways in particle beds.

## II. Experimental and Modeling Methods

### A. Characterization and Experimental Methods

Amorphous silica microspheres (Cospheric LLC, inc) with three different average particles sizes (23 μm, 67 μm and 330 μm) were chosen as the model system due to the near monodispersed nature of the particles and the well-known properties of amorphous silica. The same microspheres were used and $k_{eff}$ in vacuum was reported by Sakatani et al.[28], but in this work we measured and modeled $k_{eff}$ of these particles under different gaseous pressures (20-760 Torr). The scanning electron microscopy (SEM) images of the particles are shown in Fig. 1(a). The particles are near-perfect spheres with relatively smooth surface at the length scale of the SEM imaging. To determine the particle size of the microspheres, the mean diameters of the particles were obtained by averaging the sizes of 100 particles in the SEM images using *ImageJ* [30,31]. As shown in Figs. 1(b) to 1(d), the particle sizes follow the Gaussian distribution with a narrow deviation (< ~10% of the mean particle diameters). The average particle sizes are measured to be 23±2.7 μm, 67±4.1 μm and 330±22.8 μm, respectively. Considering the small deviation of the particle size, it is reasonable to treat these particles as monodispersed microspheres in the modeling. This can be proven by the simulation of poly-dispersed particles with real particle size distributions as shown in the Supplementary Information Section S4. The simulated $k_{eff}$ for the mono-dispersed particles (based on the average particle size) and the poly-dispersed particles (based on the size distribution shown in Fig. 1) are close. Therefore, to reduce the computational complexity, we modeled the particles assuming mono-dispersed size. The particles were also examined by the X-ray diffraction (XRD) as shown in Fig. 1(e). Evidently, these particles are amorphous silica as indicated by the diffuse and broad XRD signal[32]. As such, the bulk thermal conductivity of amorphous $SiO_2$ is used (e.g., ~ 1.3 W m$^{-1}$ K$^{-1}$ at room temperature and is temperature dependent) [33] for the modeling.

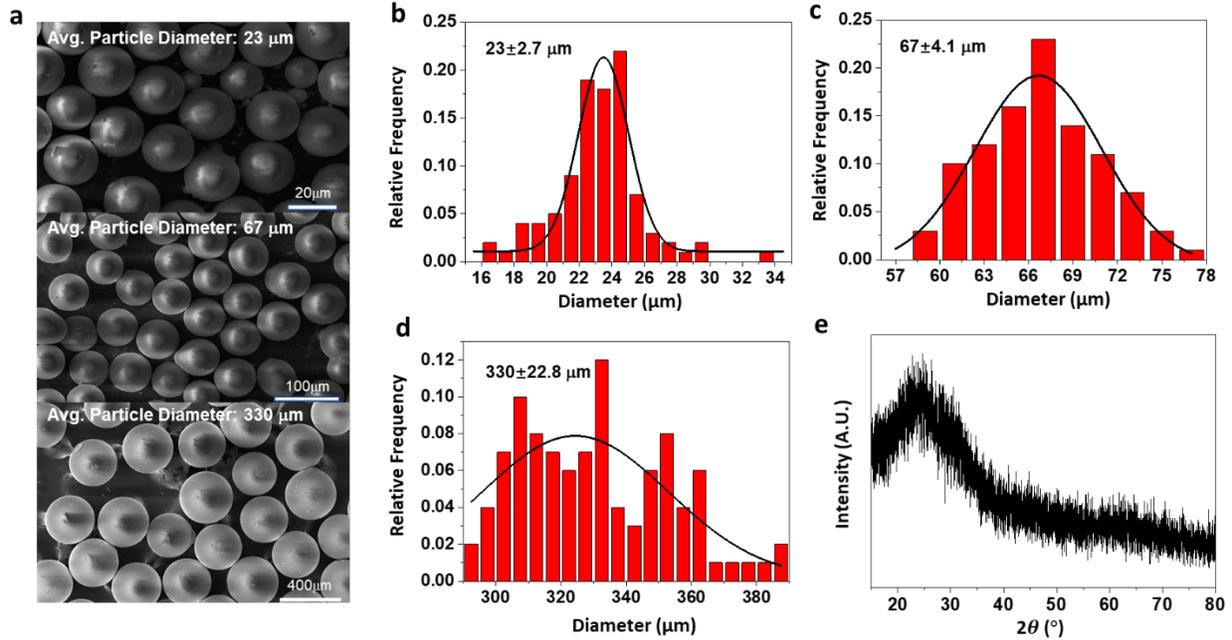

Fig. 1. Characterization of SiO$_2$ microspheres. (a) SEM images of SiO$_2$ microspheres. Particle size distribution of SiO$_2$ microspheres with average diameter of (b) 23 μm, (c) 67 μm and (d) 330 μm. 100 particles are selected from the SEM images for each size and the diameter of each particle is measured from the top view. (e) XRD pattern of SiO$_2$ microspheres, indicating amorphous phase of the silica particles.

The effective thermal conductivity $k_{eff}$ of the particle beds was measured using a THW apparatus as reported in our prior work[26]. The volume of the cavity of the THW holder holding the particles is 11.6 × 1.5 × 0.97 cm$^3$. To systematically measure $k_{eff}$ at different temperature up to 500°C and gaseous pressure from 20 to 760 Torr, the THW test section was placed in a tube furnace equipped with a N$_2$ gas supply and a pressure control system. In the THW measurement, the experimental effective thermal conductivity ($k_{eff,exp}$) of the particle bed was obtained from the transient temperature rise after applying a constant electrical current $I$ to the hot wire with the assumption of a line heat source in an infinite medium:

$$k_{eff,exp} = \frac{q'/4\pi}{d(\Delta T)/d(\ln t)} \tag{1}$$

where $\Delta T$ is the temperature rise of the Pt wire induced by applying an electrical current ($I$) to it for a duration of time ($t$), $q'$ is the heat rate per unit length of the Pt wire: $q' = I^2 R/L$, where $L$ is the length of the Pt wire and $R$ is the resistance of the Pt wire at the measurement temperature ($T$). More details of the THW system and measurement can be found in the literature[9] and in our earlier work[26].

## B. DEM Modeling

We carried out DEM simulation using the LIGGGHTS package [LAMMPS (Large-scale Atomic/Molecular Massively Parallel Simulator) Improved for General Granular and Granular Heat Transfer Simulations][34]. To obtain the real packing structure of the particle beds, a rectangular container with the dimension of $15D \times 15D \times 20D$ ($L \times W \times H$) was chosen as the simulation domain, where $D$ is the particle diameter, as shown in Fig. 2(a). The dimension was chosen in order to reduce the computational complexity while avoiding the near-wall effect [35-37]. The simulation domain was empty before the simulation as shown in Fig. 2(a) at $t = 0$, with the particle injected from the top surface of the domain set with the dimension of $15D \times 15D$ ($L \times W$). To randomly generate particles in the insertion surface, several prime numbers larger than 10000 were selected as seeds. In the simulation, the particles were generated with an initial downward velocity ($v_0$) of ~$0.1\ m\ s^{-1}$ and a global downward gravity $g = 9.81\ m\ s^{-2}$ as shown in Fig. 2(a) at $t_1 = 3.4$ ms. The bottom of the domain served as the stopping surface and thus the particles were piled up from bottom to the top in a random fashion as shown in Fig. 2(a) at $t_2 = 26.3$ ms. The simulation stopped when the entire domain was filled with particles, denoted as the equilibrium state in Fig. 2(a).

The particle properties for the simulations were based on the real particles (silica) used in our experimental measurements. The density $\rho = 2600\ kg\ m^{-3}$ indicates the particle inertia. The Young's

Modulus $E = 72\ GPa$, the Poisson's Coefficient $v = 0.278$, the Restitution Coefficient $C_R = 0.7$ and the Friction Coefficient $C_F = 0.3$ represente the inter-particulate contact behavior [38]. In addition, the cohesive energy density $e_{cohesive}$, implying the amount of energy needed to separate unit volume of particles, is applied to reflect the adhesive force between the particles at the interface [39,40]. We assumed that the cohesive energy density has the form of $e_{cohesive} = a \times D^{-b}$ and the parameter pair (a, b) is adjusted to get the optimum values of $a = 0.25$ and $b = 2$ so that the modeled particle packing density is close to the experimental value while the physical meaning of these parameters is yet to be explored. Therefore, this property was optimized so that the simulated packing densities for all the three particles were close to the experimental values. The DEM simulation parameters are listed in Table 1. Thermal conductivities of bulk silica and N₂ gas are shown in Fig. S3. The Hertz contact model was implemented for modeling the contact force. This contact force can be divided into normal and tangential components as shown in Fig. 2(b).

$$F_{contact} = F_{normal} + F_{tangential} \tag{2}$$

The normal force includes a spring force and a damping force, while the tangential force includes a shear force and a damping force. Each force was computed from the amount of overlap and the relative velocity between particles.

$$F_{normal} = F_{spring} + F_{damping,n} \tag{3}$$

$$F_{tangential} = F_{shear} + F_{damping,t} \tag{4}$$

$F_{shear}$ in Eq. (4) is proportional to the tangential displacement at the contact point, which can be calculated by integrating the relative velocity in the tangential direction over time. To take the shear force into account, we selected *tangential history* in the LIGGGHTS simulation. In addition, a simplified Johnson-Kendall-Robert (SJKR) model is used to include an additional normal force, which tends to keep particles in contact. With the SJKR model and the definition of cohesive energy density, we simulated the cohesive effect. To ensure the accuracy, we defined the neighboring distance as the radius of the particle and set the time step less than 10% of the Rayleigh/Hertz time in our simulation. Once the simulation finished (i.e., all the particles were stable), the coordinates and velocity of each particle were saved as .txt files to extract the packing structure of the particle bed, which was then imported into a COMSOL model for FEM simulation of $k_{eff}$.

Figs. 2(c) to 2(e) show the packing structure of the loosely packed silica microspheres in the *x*, *y* and *z* directions, respectively. The normalized packing density is plotted as a function of the normalized distance by the particle diameter (i.e., *L/D*, where *L* is the real distance). It is noted that Figs 2(c) and 2(d) show the similar packing structures as the particle bed is symmetric in the lateral directions while Fig. 2(e) shows packing structure in the vertical direction, with *z* = 0 as the bottom surface. As shown in Figs. 2(c) and 2(d), the near-wall effect exists within two particle sizes away from the wall, where the packing density oscillates, which agrees with the observations in the literature [35-37] as shown in the supplementary information section S1. The packing density stabilizes in the bulk particle bed (i.e., more than two particles away from the wall). It is seen that the packing density decreases with the particle size increasing from 23 μm to 330 μm. As shown in Fig. 2(e), there is also near-wall effect close to the bottom surface within two particles away from the wall. Except that the packing density decreases close to the top surface because this surface is unconstrainted with continuous inflow of particles, the packing structure in the *z* direction is similar to that in the *x* and *y* directions, indicating that the packing structure is homogeneous in the bulk. As shown in Table 2, the average bulk packing density is 0.57, 0.55 and 0.46 for 23 μm, 67 μm and 330 μm particles, respectively, which is closed to that measured in the experiment. The theoretical packing density of a simple cubic structure is calculated to be 0.52 and is invariant with particle size. Evidently, the simple cubic structure does not represent the actual packing of real particle beds.

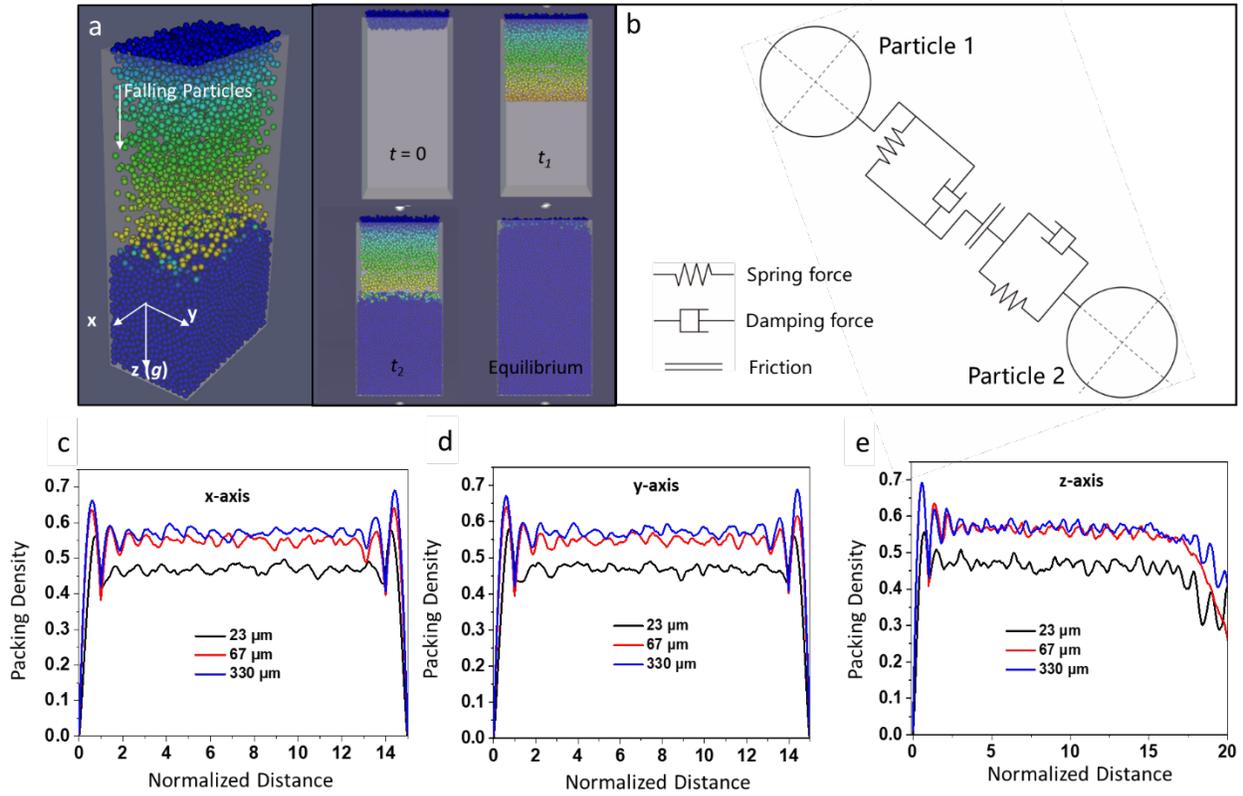

Fig. 2. DEM simulation of packing structure. (a) Snapshots of randomly packed $SiO_2$ microspheres at different time. The packing structure is extracted from the equilibrium state. (b) Schematic of the forces between two adjacent particles in DEM. Distribution of packing density at (c) *x* direction, (d) *y* direction, and (e) *z* direction. The packing density fluctuates near the wall and is a constant in the middle of the particle bed where the packing density is extracted.

Table 1. Simulation Parameters for DEM modeling

| Parameters | Values |
|---|---|
| Particle Diameter | $D$ |
| Skin Distance | $D/2$ |
| Time Step | $2.5 \times 10^{-10}\ [sec]$ |
| Young's Modulus | $72\ [Gpa]$ |
| Poisson's Ratio | 0.278 |
| Coeff. Of Restitution | 0.7 |
| Coeff. Of Friction | 0.3 |
| Cohesion Energy Density | $0.25 \times D^{-2}\ [J\ m^{-3}]$ |
| Particle Density | $2600\ [kg\ m^{-3}]$ |
| Gravity | $9.81\ [m\ s^{-2}]$ |
| Initial Velocity | 0.1 $[m\ s^{-1}]$ |

Table 2. Comparison of average packing density of SiO$_2$ particle bed.

| Packing Structure | Packing Density (%) | | |
|---|---|---|---|
| | $D$ = 330 μm | $D$ = 67 μm | $D$ = 23 μm |
| Experiment | 0.57 | 0.55 | 0.43 |
| Random (DEM) | 0.57 | 0.55 | 0.46 |
| Simple Cubic (SC) | 0.52 | 0.52 | 0.52 |

## C．Modeling of Effective Thermal Conductivity

Effective thermal conductivity of particle beds is contributed by gas conduction, solid conduction via contacts between particles, and radiation heat transfer. The modeled effective thermal conductivity ($k_{eff,calc}$) consists of two parts: solid and gas thermal conductivity ($k_{s,g}$) modeled using COMSOL and radiation thermal conductivity ($k_r$) modeled analytically based on the optical properties of silica.

$$k_{eff,calc} = k_{s,g} + k_r \tag{5}$$

The coordinates of all the particles were extracted from the DEM simulation as the input into the COMSOL model to compute $k_{s,g}$. As shown in Fig. 3(a), a $4D \times 4D \times 4D$ ($L \times W \times H$) cube was chosen as the COMSOL simulation domain, which was truncated from the middle of the particle bed with a distance of at least two particles away from the walls to avoid the near-wall effect. Since the packing structure (away from the walls) is homogeneous as shown in Fig. 2(c,d,e), the simulation result is insensitive to the position of the chosen domain, which is also proven by the grid independence test in |Supplementary Information Section S2. The effect of different packing methods was also tested in this study, e.g., a shaking process with 10g acceleration was added after all particles were settle down in the previous DEM simulation. As shown in Supplementary Information Section S5, the results remained the same, indicating the independence of the random packing structure on the packing method. The contact area between the particles was automatically given by the particle coordinates data from the DEM simulation based on the JKR theory[41] using the mechanical properties of SiO$_2$. The gas gap between the particles is filled with N$_2$ with the Kapitza resistance ($R_K$) applied to the gas-solid interface:

$$R_K = \frac{2-\alpha}{\alpha} \frac{2}{\gamma+1} \frac{L}{\mu C_v} \tag{6}$$

where $\gamma$ is the specific heat ratio of the N$_2$ gas ($\gamma$ =1.4), $\mu$ is the viscosity of the gas, $C_v$ is the specific heat of the gas at a constant volume and $L$ is the mean free path of the gas. In Eq. (6), the thermal accommodation coefficient (TAC) $\alpha$ has an important effect on the thermal resistance across a gas gap between particles. A correlation of TAC for engineering surfaces[42] was implemented:

$$\alpha = e^{-0.57\left(\frac{T}{273}-1\right)} \frac{1.4 M_g}{6.8 + 1.4 M_g} + \left[1 - e^{-0.57\left(\frac{T}{273}-1\right)}\right] \frac{2.4\mu^*}{(1+\mu^*)^2} \tag{7}$$

where $T$ is the temperature of the solid, $M_g$ is the molecular weight for diatomic gases and $\mu^*$ is the ratio of molecular weight of the gas and the solid. For comparison, a $4D \times 4D \times 4D$ cube based on the simple cubic (SC) packing of silica microspheres was also established as shown in Fig. 3(b).

For the COMSOL modeling on both random packing and SC structures, the bottom surface was set at the ambient temperature $T_0$ and the top surface was set at $T_1 = T_0 + 1K$ to create a temperature gradient along the vertical direction. Other boundaries were set as symmetric (adiabatic) boundaries. The heat flux $q_s$ was extracted from the bottom surface from the normal heat flux of $4D \times 4D$ area, given by COMSOL and the conduction thermal conductivity is calculated by:

$$k_{s,g} = \frac{H q_s}{T_1 - T_0} \tag{8}$$

where $H$ is the height of the simulation domain ($H = 4D$).

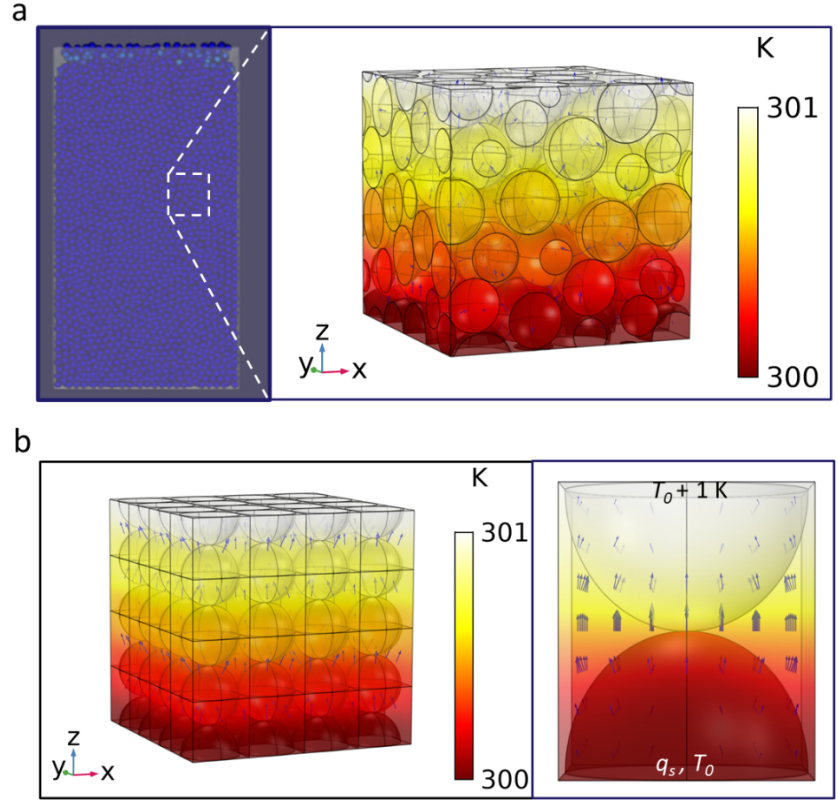

Fig. 3. COMSOL simulation of thermal conductivity of SiO₂ particle bed based on (a) random packing structure obtained from the DEM simulation and (b) simple cubic (SC) structure. The bottom surface is set at $T_0$ and the top surface at $T_0 + 1\ [K]$ to create a temperature gradient. The heat flux is extracted from the bottom surface.

The radiative thermal conductivity $k_r$ is modeled based on the solid-to-solid radiative heat transfer between two particles which is valid for opaque materials in the IR spectrum of interest ($D_p \leq \sim 600$ μm) [11,15,43]:

$$k_r = \frac{4\sigma}{(\frac{2}{\varepsilon_r} - 1)} T_0^3 \quad (9)$$

where $\sigma$ is the Stephan-Boltzmann constant, and $\varepsilon_r$ is the thermal emittance of silica quoted from Ref.[44], as shown in Supplementary Information Section S3. This model has been proven valid for black ceramic particles with the particle sizes < ~ 600 μm[11,43]. With the $k_{s,g}$ and $k_r$, the total effective thermal conductivity $k_{eff,calc}$ was then calculated using Eq. (5).

## III. Results and Discussion

**Figs. 4(a) to 4(c)** show the experimental results of silica microsphere beds with the particle sizes of 23 μm, 67 μm and 330 μm, respectively, up to 500 °C under the $N_2$ gas environment with the gaseous pressure from 20 to 760 Torr. The maximum temperature was set to avoid the phase change and sintering of silica microspheres. For all the particles at 760 Torr, $k_{eff}$ increases with increasing temperature mainly due to the enhanced gas conduction, a slight increase in the solid conductivity and higher $k_r$ at higher temperature. It is noted that the radiation contribution is much less significant compared to the gas conduction within the temperature range in our measurement, which is shown by the fact the $k_{eff,exp}$ at 20 Torr is independent of temperature for the 23 and 67 μm particles and only slightly dependent on the temperature for the 330 μm particles because at this gas pressure, the gas conduction is suppressed. Previous studies indicate that the characteristic length scale for gas conduction $L_g \approx \frac{D}{7}$ instead of $D$ because the gas bridge close to the contact point dominates the gas conduction[26]. Therefore, $L_g$ is ~ 3.3 μm, ~ 9.6 μm and ~ 47 μm for the 23 μm, 67 μm and 330 μm particles, respectively. For the 23 μm microspheres, the gas conduction is completely suppressed at 20 Torr because the mean free path of $N_2$ at 20 Torr and room temperature is 2.3 μm, comparable to $L_g$. Therefore, it can be concluded that the contribution of the solid conduction to total $k_{eff}$ for the 23 μm particles is less than 0.02 W m$^{-1}$ K$^{-1}$. For the 67 μm and 330 μm particles, the gas conduction is not completely suppressed at 20 Torr as the mean free path of $N_2$ at 20 Torr is still smaller than $L_g$, so $k_{eff}$ at this pressure is higher than that of the 23 μm particles.

The modeled effective thermal conductivity ($k_{eff,calc}$) based on the simple cubic structure and the random packing structure by DEM is compared to the experimental results in Figs. 4(a) to 4(c). As shown in Fig. 4(a) for the 23 μm particles, the modeling result based on the simple cubic structure is higher than the experimental one at all the temperatures and gas pressures while the results based on the random packing structure agree better with the experimental results. The reason is that the simple cubic structure overestimates the conduction heat transfer via the solid-solid pathway, caused by the overestimation of the particle packing density as shown in Table 2. This is clearly shown in the overestimation of $k_{eff}$ at 20 Torr where the solid conduction $k_s$ become relatively more important than the gas conduction $k_g$. As shown in Fig. 4(b) for the 67 μm particles, both the simple cubic and random packing models capture the experimental results well at 100 Torr and higher pressure because both models predict the real packing density of the particles of this size as shown in Table 2. However, the simple cubic model still has the same overestimation of solid conduction and hence a larger relative error at 20 Torr. As shown in Fig. 4(c) for the largest particle size (330 μm), contrary to that for the 23 μm particles, the model based on the simple cubic structure underestimates $k_{eff}$ while the model based on the random structure agree well with the experimental results because the simple cubic structure underestimates the packing density of the large particles as shown in Table 2.

Figs. 4(d) to 4(f) compare $k_{eff}$ as a function of gas pressure of the three particle beds at 27°C, 220°C and 420°C respectively. It is seen that the random packing model captures the trend well for all the particle size and temperatures. Meanwhile, $k_{eff,calc}$ from simple cubic model deviates drastically from the experimental results in almost all the cases because the gas conduction is the major heat pathway in particle beds, which strongly relies on the accurate packing structure.

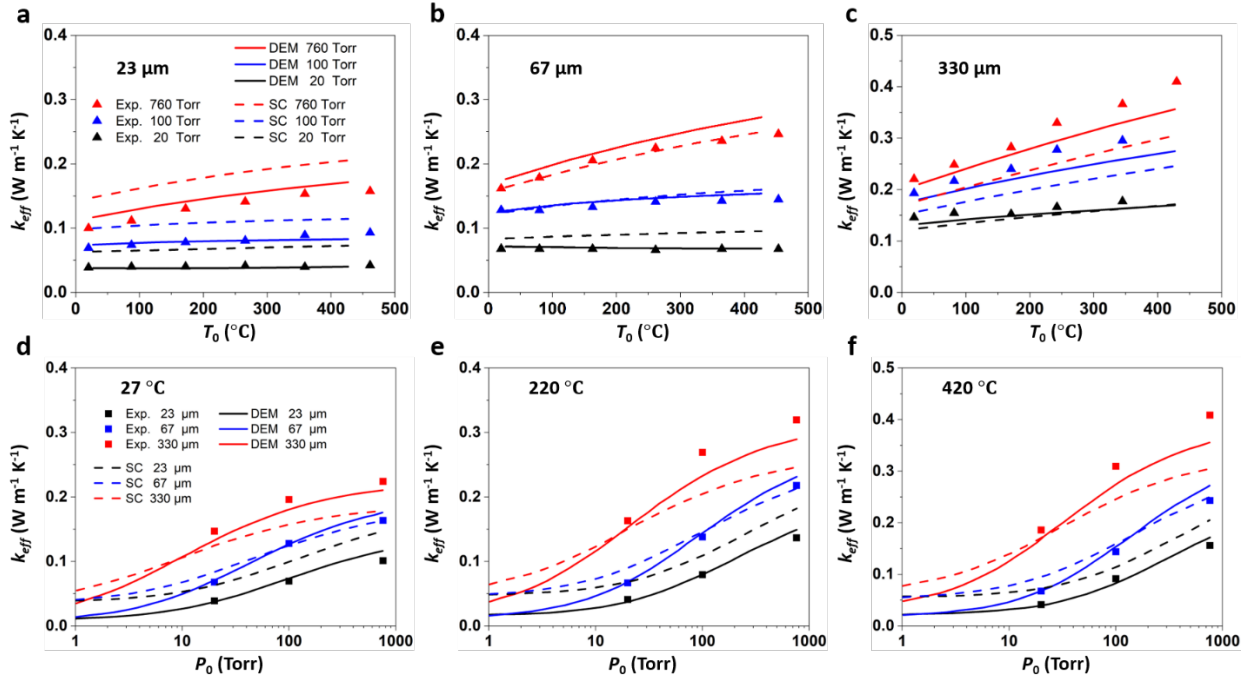

Fig. 4. Comparison of experimental and simulated thermal conductivity of $SiO_2$ based on simple cubic (SC) packing and DEM random packing. Thermal conductivity of $SiO_2$ particle bed as a function of temperature at different gaseous pressures for the diameter of (a) 23 μm, (b) 67 μm and (c) 330 μm particles for gas pressure at 760 Torr, 100 Torr and 20 Torr. Thermal conductivity of $SiO_2$ with different particle sizes as a function of $N_2$ gaseous pressure at (d) 27C°, (e) 220C° and (f) 420C° for particle sizes of 23 μm, 67 μm and 330 μm. Solid lines represent $k_{eff,calc}$ based on the random packing structure extracted from DEM; Dashed lines represent $k_{eff,calc}$ based on the SC packing structure; Symbols represent $k_{eff,exp}$.

To quantify the deviation of the simulated thermal conductivity from the experimental values based on the simple cubic and DEM random packing structures, we define the relative deviation $\Delta_k$ as

$$\Delta_k = \left| \frac{k_{eff,calc} - k_{eff,exp}}{k_{eff,exp}} \right| \tag{10}$$

where $k_{eff,exp}$ is the experimentally measured thermal conductivity (Eq. (1)) and $k_{eff,calc}$ is the modeled thermal conductivity (Eq. (5)). Figs. 5(a) to 5(c) show the $\Delta_k$ for 23 μm, 67 μm and 330 μm microspheres, respectively at various temperatures and pressures. It is evident that $\Delta_k$ for the DEM model is significantly smaller than that of the simple cubic model in all the cases. For the 23 μm microspheres as shown in Fig. 5(a), $\Delta_k$ is within 15% for the DEM model at all the temperatures and gas pressures while that of the simple cubic model is greater than 30% in most cases. For the 67 μm microspheres, $\Delta_k$ for the simple cubic model is < 11% at 760 and 100 Torr but reaches 41% at 20 Torr while the error of the DEM model is less than 12% for all the pressures. The DEM model can also capture the thermal conductivity of the 330 μm microspheres with $\Delta_k$ less than 13% as shown in Fig. 5(c). Besides, it can be noticed that $\Delta_k$ of the simple cubic model is no greater than 25% for the 330 μm microspheres, which is better than that for smaller particles. It has been reported that the unit cell models (e.g., the ZBS model) could fit the thermal conductivity of large particles (> 300 μm) well[11,26] because the contribution from solid conduction becomes less significant for larger particles. Based on the analysis above, we can conclude that the DEM model agrees significantly better with the experimental results than the simple cubic model due to the realistic packing structure.

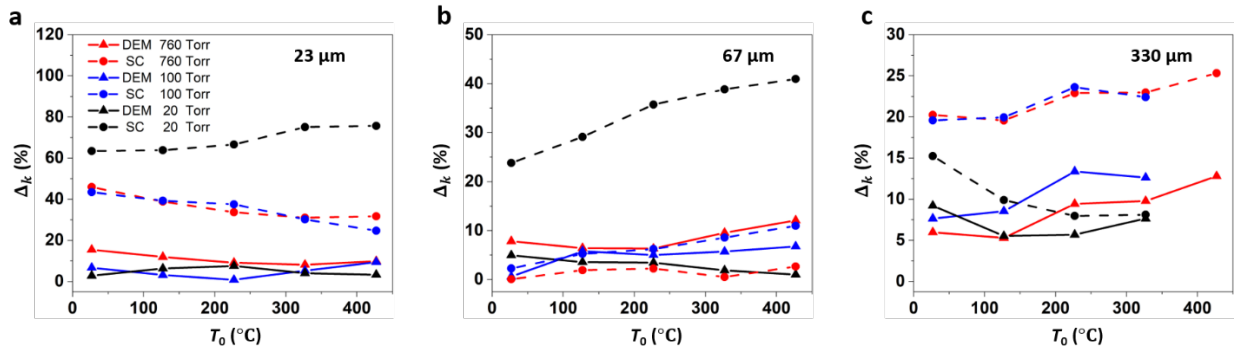

Fig. 5. Comparison of deviation from the experimental value based on simple cubic (SC) and DEM packing structures at different temperature and gas pressure for (a) 23 μm, (b) 67 μm and (c) 330 μm particles at different gas pressures. The relative deviation is defined as the percentage difference between the modeled and measured thermal conductivity values.

The analysis demonstrates that the random packing structure obtained from the DEM is better for the prediction of the effective thermal conductivity of particle beds compared to the simple cubic structure that was commonly used in the literature. Using this model, the contributions from different heat transfer modes can be more accurately quantified, including the gas conduction in the void between the particles, solid conduction through the particle-particle contact, and radiative heat transfer. As shown in Eq. (5), $k_{eff,calc}$ is determined from the combined solid and gas conduction components ($k_{s,g}$) obtained from the COMSOL simulation and the radiative component $k_r$. To further differentiate the gas conduction $k_g$ and the solid conduction $k_s$ components from $k_{s,g}$, COMSOL simulation on the random packing structure from DEM model was conducted under vacuum condition (i.e., gaseous pressure is set at 0.01 Torr to mimic the vacuum condition to avoid singularity problem in the COMSOL modeling, which is valid due to the high Kapitza resistance at 0.01 Torr compared to the resistance in the bulk gas), where the gas conduction is eliminated and thus $k_{s,g}$(vacuum) =$k_s$(at $P$ = 20, 100, 760 Torr). Therefore, we can also obtain $k_g$ at 760 Torr by subtracting $k_s$ from $k_{s,g}$ at the same pressure. Figs. 6(a) to 6(c) show the contributions of $k_s$, $k_g$ and $k_r$ to $k_{eff,calc}$ under 760 Torr $N_2$ for 23 μm, 67 μm and 330 μm silica microsphere beds, respectively. The gas conduction is the dominant heat transfer mechanism for all the three particle sizes. For example, at $T_0 = 700°C$, $k_g$ contributes to ~ 90% of $k_{eff,calc}$ to all the three sizes. The solid conduction component $k_s$ for the 23, 67 and 330 μm $SiO_2$ microspheres is calculated to be similar (~ 0.007 W m$^{-1}$ K$^{-1}$) at room temperature. Because of the increasing total $k_{eff}$ with particle size, the relative contribution of $k_s$ becomes less important with increasing size. Measured $k_s$ under vacuum of similar $SiO_2$ microspheres in Ref [28] also showed a near constant, albeit lower, value of 0.002 to 0.003 W m$^{-1}$ K$^{-1}$ for the particle size range of 50-500 μm. The higher $k_s$ value obtained from our model could be due to the simplification of size distribution and particle morphology used in the DEM model. Specifically, it was found that the $SiO_2$ microspheres have μm-scale roughness and surface contaminants, which are expected to reduce the $k_s$ due to the solid contacts[28], whereas in our DEM model the particles are assumed to be perfectly spherical and smooth. As such, a higher $k_s$ value is expected from our DEM model compared to the experimental data on the real microspheres in Ref.[28].

The model allows us to quantify the relative contributions from each heat transfer mechanisms for different packing structures and different particle sizes. The relative contribution of $k_s$ becomes less important for the larger particles. For example, based on the DEM packing structure, at $T_0 = 700°C$, $k_s$ contributes to ~ 13.9% of $k_{eff,calc}$ for the 23 μm particles, which decreases to ~ 4.4% for the 330 μm particles. On the contrary, the radiation contribution, $k_r$, increases with increasing particle size due to the longer propagation length of photons, e.g., at $T_0 = 700°C$, $k_r$ only contributes to ~ 1.2% of $k_{eff,calc}$ for the 23 μm particles but it increases to ~ 7.5% for the 330 μm particles. To investigate the effect of packing structure on the heat transfer mechanisms, the gas conduction $k_g$ and solid conduction $k_s$ components from $k_{s,g}$ for the simple cubic structure are also differentiated in the same way as for the DEM model and plotted in Fig. 6(a) to 6(c). Notably, the contribution of $k_s$ is overestimated when using the simple cubic structure, e.g., $T_0 = 700°C$, $k_s$ contributes to ~ 27.8% of $k_{eff,calc}$ for the 23 μm particles and ~ 15.2% for the 330 μm particles. On the contrary, the contribution of $k_g$ is underestimated by the simple cubic structure, e.g., $T_0 = 700°C$, $k_g$ contributes to only ~ 75% of $k_{eff}$ for the simple cubic structure vs. ~ 90% for the random packing structure. A possible reason is that in the random packing structure, the number of gas bridges between the particles (i.e., the region ~ 1/6 to 1/12 particle diameter away from the contact point[45,46]) increases, which reduces the effective length of gas conduction pathway. Although the theoretical coordination number of simple cubic structure is six, the effective coordination number for heat transfer should be considered as two because of the near one-dimensional heat transfer as shown in Fig. 3(b). The coordination number in the DEM for 23 μm, 67 μm and 330 μm diameter particle bed is calculated to be 4.2, 3.5 and 3.1, respectively, which should be considered as the effective coordination number due to the three-dimensional heat transfer in the random packing. The higher effective coordination number in random packing structure may explain the higher relative contribution from the gas conduction. Nevertheless, the effective length of solid conduction may be increased in the random packing structure because of the serpentine solid conduction pathway compared to the straight one in the simple cubic structure, which reduces the relative contribution from the solid conduction. The results above indicate that to accurately understand the heat transfer mechanism in a particle bed, it is imperative to know the real packing structure.

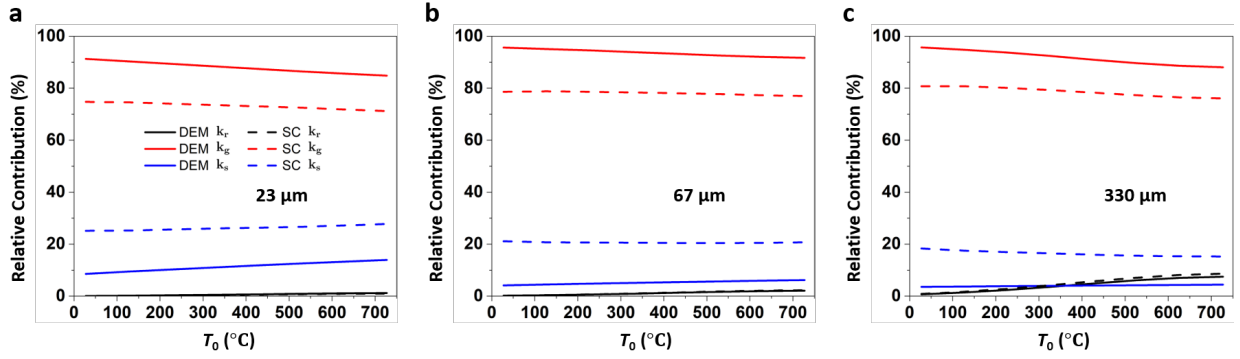

Fig. 6. Relative contribution of different heat transfer pathways (radiative heat transfer $k_r$, gas conduction $k_g$ and solid conduction $k_s$) to the effective thermal conductivity of SiO$_2$ particle beds calculated from Eq. (5) for (a) 23 μm, (b) 67 μm, and (c) 330 μm SiO$_2$ microspheres based on the simple cubic (SC) structure and the random packing structure from DEM (DEM).

## IV. Conclusion

We measured and modeled the effective thermal conductivity of silica particle beds made of near monodispersed microspheres. $k_{eff,exp}$ of the particle beds was measured for particles with average diameter of 23, 67, and 330 μm and with varying N$_2$ gaseous pressures from 20 to 760 Torr and temperatures up to 500 °C to quantify the relative contributions of solid conduction, gas conduction and radiation heat transfer in the particle beds. To better understand the effect of the particle packing structure on $k_{eff}$, modeling based on the discrete element method (DEM) was employed using the LIGGGHTS package to simulate the particle packing process. It was found that the random structure obtained from the DEM simulation better represents the real packing structure in the particle beds compared to the simple cubic structure normally assumed in common heat transfer models for particle beds, as demonstrated by the good agreement of packing density between the experimental and DEM simulation results. The random packing structure extracted from the DEM was then imported into a COMSOL-based FEM model to simulate the effective thermal conductivity ($k_{eff,calc}$) of the particle beds. $k_{eff,calc}$ based on the random packing agree better with the experimental results for all the three particle beds with various gas pressures and temperatures. Compared to the simple cubic model, the random packing model can better capture $k_{eff}$, indicating the importance of the actual packing structure of particles for heat transfer in particle beds. Based on the simulation results from the random packing, we found that gas conduction contributes to ~ 90% of the $k_{eff}$ for the particle beds at 760 Torr while solid conduction and radiation heat transfer only contribute to ~ 10%. In contrast, the model based on the simple cubic structure predicts a lower contribution from gas conduction (~ 75%) while higher contribution from solid conduction possibly due to the longer effective length of gas conduction pathway in the simple cubic structure. This indicates that the packing structure impacts both the solid and gaseous conduction pathways. Additionally, the model confirms that the relative solid conduction is found to be more important for smaller particles while radiation heat transfer increases with larger particles due to the longer propagation length of photons.

**Acknowledgements**

This material is based upon work supported by the U.S. Department of Energy's Office of Energy Efficiency and Renewable Energy (EERE) under Solar Energy Technologies Office (SETO) Agreement Number DE-EE0008379. The views expressed herein do not necessarily represent the views of the U.S. Department of Energy or the United States Government.